\documentclass[runningheads]{llncs}

\usepackage{enumitem}
\usepackage{booktabs}
\usepackage{amsmath} %
\usepackage[T1]{fontenc}
\usepackage{graphicx}
\usepackage[skip=4pt]{caption}       %
\setlist{nosep}
\usepackage{etoolbox}
\usepackage{subcaption}
\usepackage{hyperref}
\usepackage{color}
\usepackage{accessibility}

\usepackage[font=footnotesize]{caption} 
\usepackage[withpage, printonlyused,nohyperlinks]{acronym}

\acrodef{hci}[HCI]{human-computer interaction}
\acrodef{geq}[GEQ]{game experience questionnaire}
\acrodef{dsp}[DSP]{digital signal processing}
\acrodef{adhd}[ADHD]{Attention Deficit Hyperactivity Disorder}

\acrodef{xr}[XR]{extended reality}
\acrodef{mr}[MR]{mixed reality}
\acrodef{ar}[AR]{augmented reality}
\acrodef{vr}[VR]{virtual reality}

\acrodef{bcmi}[BCMI]{brain-computer music interface}
\acrodef{bci}[BCI]{brain-computer interface}
\acrodef{mi}[MI]{motor imagery}
\acrodef{erp}[ERP]{event-related potential}
\acrodef{errp}[ErrP]{error-related potentials}
\acrodef{eeg}[EEG]{electroencephalography}
\acrodef{lda}[LDA]{Linear Discriminant Analysis}
\acrodef{svm}[SVM]{Support Vector Machine}
\acrodef{vep}[VEP]{visually evoked potentials}
\acrodef{aerp}[aERP]{auditory event-related potential}
\acrodef{ssvep}[SSVEP]{steady state visually evoked potentials}
\acrodef{tvep}[tVEP]{time-modulated visually evoked potentials}
\acrodef{mvep}[mVEP]{motion-onset visually evoked potentials}
\acrodef{cvep}[cVEP]{code-modulated visually evoked potentials}
\acrodef{fft}[FFT]{Fast Fourier Transform}
\acrodef{psd}[PSD]{Power Spectral Density}
\acrodef{csp}[CSP]{Common Spatial Patterns}
\acrodef{ers}[ERS]{Event-Related Synchronization}
\acrodef{erd}[ERD]{Event-Related Desynchronization}
\acrodef{snr}[SNR]{signal-to-noise ratio}
\acrodef{mdrm}[MDRM]{Minimum Distance to Riemannian Mean}
\acrodef{smr}[SMR]{sensory motor rhythms}
\acrodef{itr}[ITR]{Information Transfer Rate}

\acrodef{ml}[ML]{machine learning}
\acrodef{ai}[AI]{Artificial Intelligence}
\acrodef{dl}[DL]{deep learning}
\acrodef{dnn}[DNN]{deep neural network}
\acrodef{cnn}[CNN]{Convolutional Neural Network}

\begin{document}
\title{BrainForm: a Serious Game for BCI Training and Data Collection}
\titlerunning{BrainForm: a Serious Game for BCI Training and Data Collection}
\author{Michele Romani\inst{1,}\inst{2}\orcidID{0009-0002-3446-819X} \and
Devis Zanoni\inst{1}\orcidID{0009-0005-7446-5227} \and
Elisabetta Farella\inst{2}\orcidID{0000-0001-9047-9868} \and Luca Turchet\inst{1}\orcidID{0000-0003-0711-8098}}
\authorrunning{M. Romani et al.}
\institute{University of Trento, 38122, Trento, Italy\and
Fondazione Bruno Kessler, 38122, Trento, Italy}
\maketitle              %
\begin{abstract}
\textit{BrainForm} is a gamified Brain-Computer Interface (BCI) training system designed for scalable data collection using consumer hardware and a minimal setup. We investigated (1) how users develop BCI control skills across repeated sessions and (2) perceptual and performance effects of two visual stimulation textures. Game Experience Questionnaire (GEQ) scores for \textit{Flow}, \textit{Positive Affect}, \textit{Competence} and \textit{Challenge} were strongly positive, indicating sustained engagement. A within-subject study with multiple runs, two task complexities, and post-session questionnaires revealed no significant performance differences between textures but increased ocular irritation over time. Online metrics—\textit{Task Accuracy}, \textit{Task Time}, and \textit{Information Transfer Rate}—improved across sessions, confirming learning effects for symbol spelling, even under pressure conditions. Our results highlight the potential of \textit{BrainForm} as a scalable, user-friendly BCI research tool and offer guidance for sustained engagement and reduced training fatigue.

\end{abstract}
\section{Introduction}

While \ac{bci} technology has advanced substantially, practical applications continue to face significant challenges. Issues like non-stationary signals and wide inter individual variability in physiological responses hinder the adoption of these interfaces. Most systems still require extensive calibration for each task, involving data collection during task performance, classifier training, and subsequent online\footnote{In the context of BCI, online is synonym of real-time} evaluation. During the calibration phase, users must focus on performing the training task properly, whilst during evaluation they must learn to modulate their brain activity in response to system feedback. Often this process must be reiterated multiple times increasing costs for researchers and fatiguing the subjects.

Serious games, already widely used for testing and simulation in AI research \cite{yannakakis_artificial_2018}, are gaining traction in neuroscience and \ac{bci} research, and are reportedly becoming appealing for large-scale data collection \cite{burelli_playing_2024}. Building on these premises, we developed \textit{BrainForm}, a serious game that employs \acp{erp} to enable players to defeat enemies and solve puzzles by focusing on flickering targets. Designed for consumer hardware, \textit{BrainForm} serves as both a training environment for novice users to acquire \ac{bci} control skills and a scalable research platform capable of recording raw \ac{eeg} signals and detailed in-game metadata.

Our primary objective was to investigate how players develop \ac{bci} control skills through repeated, game-based interaction, focusing on the co-evolution of neuro-physiological adaptation and game-play mastery. To support this goal, we implemented a structured protocol consisting of an introductory tutorial, two practice runs involving calibration and control with up to ten flickering targets, and a final timed challenge designed to assess performance under pressure. Participants also completed optional free-play runs and post-session questionnaires.

In parallel, we assessed visual fatigue and perceptual comfort by comparing two stimulation textures. Since \ac{erp}-based \acp{bci} rely on repetitive flickering patterns that may cause ocular strain, thus degratating performance \cite{BLANCODIAZ2022109722}, we evaluated a standard checkerboard texture against a grain-like, randomly distributed pattern of comparable spatial density. The latter was designed to reduce visual strain while maintaining stimulation effectiveness, and to test the feasibility of integrating \ac{bci} visual elements into game aesthetics.

Each participant experienced both visual texture conditions in a counterbalanced order. We addressed the following research questions:
{\setlength{\itemsep}{0pt}
\setlength{\parskip}{0pt}
\setlength{\topsep}{0pt}
\begin{itemize}[leftmargin=1em]
\item[•] \textbf{RQ1:} \textit{How does the design of visual stimulation patterns (e.g., texture type) impact players’ perceived visual fatigue in ERP-BCI?}
\item[•] \textbf{RQ2:} \textit{Does visual texture design influence the effectiveness of ERP-based \ac{bci} control, as reflected in user performance?}
\item[•] \textbf{RQ3:} \textit{How do players' \ac{bci} control skills evolve over repeated game-play runs?}
\item[•] \textbf{RQ4:} \textit{How does a pressure element impact players’ \ac{bci} control performance?}
\end{itemize}
}

By combining controlled experimentation with engaging game-play, our study makes three main contributions:
(1) we introduce \textit{BrainForm}, a gamified \ac{bci} platform designed for scalable data collection using consumer-grade hardware;
(2) we present a structured, game-based training protocol that enables simultaneous assessment of \ac{bci} control skill acquisition and perceptual comfort; and
(3) we provide empirical insights into how visual stimulation design and gameplay pressure influence user performance and fatigue in ERP-based \acp{bci}.

The remainder of this paper is organized as follows:  Section~\ref{background} introduces the types of \ac{bci} and relevant metrics; 
Section~\ref{related_work} delves into the foundational and related works;
Section~\ref{methods} describes the experimental setup and study protocol; Section~\ref{results} reports the behavioral, performance, and perceptual outcomes; Section~\ref{discussion} discusses implications for gamified \ac{bci} design; and Section~\ref{conclusions} concludes with directions for future work.

\begin{figure}[h]
    \centering
    \includegraphics[width=1.0\textwidth]{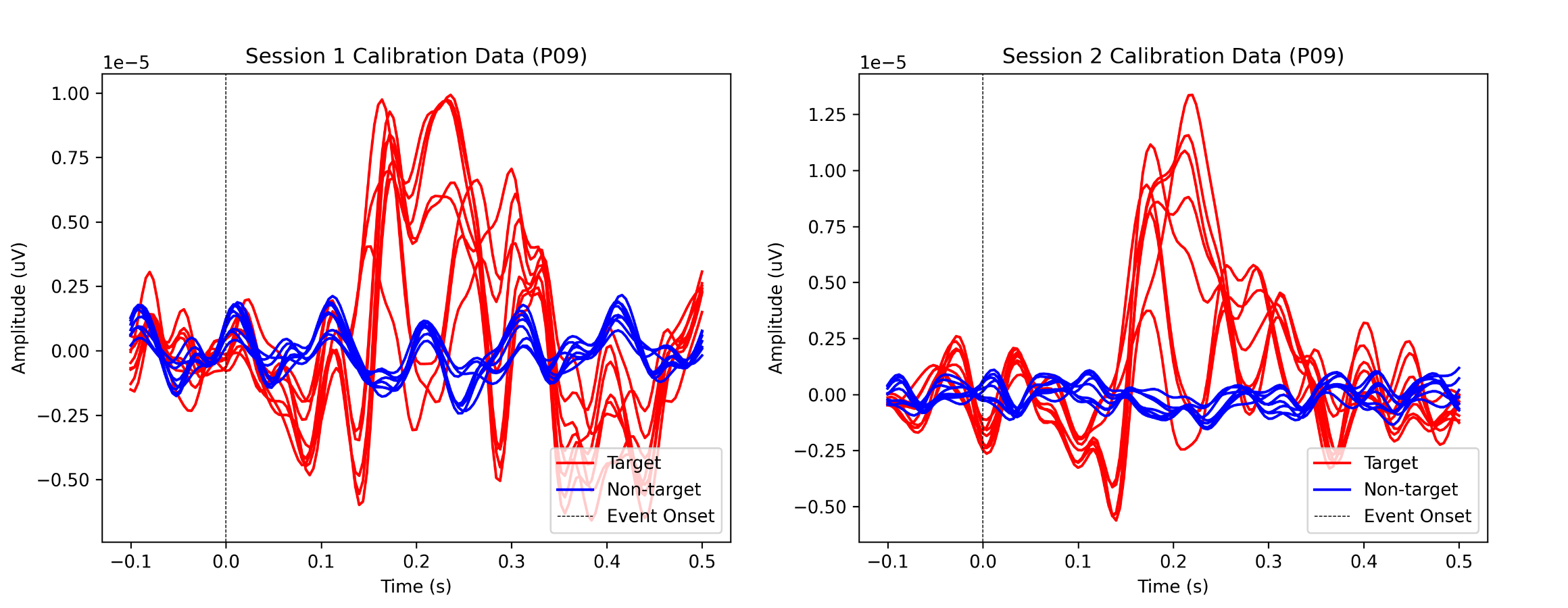}
    \caption{An example of ERPs from the calibration sessions of a subject playing \textit{BrainForm}. Calibrations trials are averaged across the session and highlighted in red:\textit{Target} and blue:\textit{Non-Target}.}
    \label{fig:erp}
\end{figure}

\section{Background}\label{background}

\subsection{Brain-Computer Interfaces}\label{types}
Most current \ac{bci} systems rely on \ac{eeg} technology, widely adopted for its good tradeoff between high temporal resolution and relatively low hardware cost. Signals are captured via scalp electrodes, using gel or dry materials like gold or polymers, and amplified before processing.
\acp{bci} are generally categorized as \textit{active}, \textit{reactive}, or \textit{passive}
\cite{zander_utilizing_2009,gurkok_braincomputer_2012}: \textit{active} \acp{bci} rely on intentional modulation, such as \ac{mi} of limb movements; \textit{reactive} \acp{bci} depend on focused attention to external stimuli, triggering \acp{erp}; \textit{passive} \acp{bci} monitor brain activity continuously to infer states like emotion or error detection, without user control.
This work focuses on reactive ERP-\acp{bci}. An example of elicited \acp{erp} is shown in Fig.\ref{fig:erp}.

\begin{figure}[h]
    \centering
    \includegraphics[width=\columnwidth]{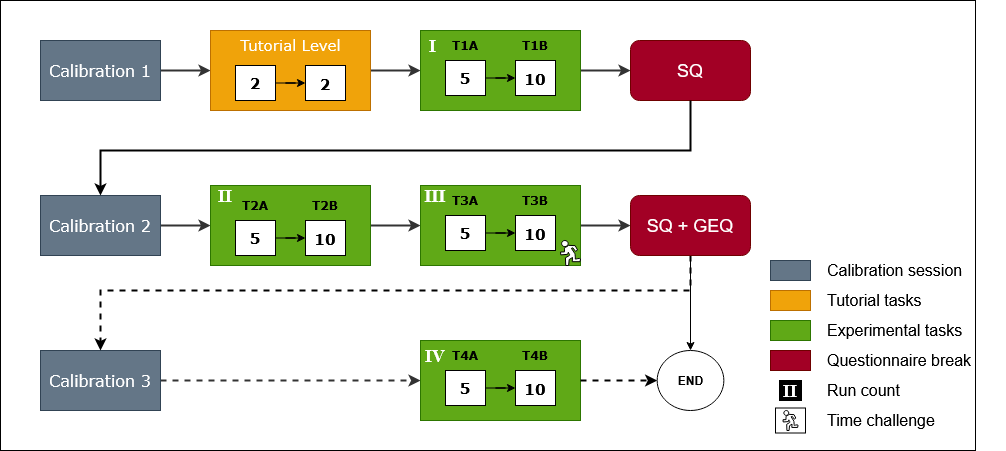}
    \caption{The experimental protocol of \textit{BrainForm}. Runs included in the statistical analysis are highlighted in green. The numbers in white boxes represents the amount of flickering targets in the task.}
    \label{fig:protocol}
\end{figure}

\subsection{Metrics}
Common metrics in the \ac{bci} research field focus on measuring the reliability and the speed of the system.
In an online classification system, preformance scores result from a combinations of classifier outputs and other mechanisms in place to prevent potential false activations (see section \ref{configuration}).
\textit{Task accuracy} defined as:
$$
    T_{\text{acc}} = \frac{C_{\text{class}}}{C_{\text{class}} + M_{\text{class}}} \eqno{(1)}
$$
where ${C_{\text{class}}}$ are the correctly classified targets and ${M_{\text{class}}}$ are the misclassified targets; the \textit{task time} is the elapsed task time from the beginning of the task:
$$
    T_{\text{time}} = T_{\text{end}} - T_{\text{start}} \eqno{(2)}
$$
Finally, the \ac{itr} measures the total throughput of the \ac{bci} in bits/s as defined by Wolpaw \cite{712231} by the following formula, accordingly adapted to the naming convention of this study:
$$
    B = \left(\frac{N_{trials}}{T_{time}}\right) * \left[log_2N + T_{acc}log_2T_{acc} + \left(1 - \frac{T_{\text{acc}}}{N-1}\right)\right]  \eqno{(3)}
$$
where the first terms equates to the time per selection, and N is the number of possible choices. Removing the first term gives the \ac{itr} in bits/symbol.

\section{Related Work}\label{related_work}

ERP-based \acp{bci} enable direct communication between the brain and external devices by detecting neural responses to time-locked stimuli, typically measured via scalp-recorded EEG. Among the most widely used ERP paradigms are the \ac{ssvep} and P300, both well-suited for fast-paced interaction scenarios \cite{kaplan_adapting_2013}. These methods rely on the brain’s consistent responses to visual stimuli or cues and have demonstrated high accuracy in applications like communication systems and speller interfaces \cite{farwell_talking_1988,lalor_steady-state_2005}.

In recent years, alternative stimulation paradigms such as \ac{cvep}, \ac{tvep}, \ac{mvep}, and \ac{aerp} have expanded the \ac{bci} design space by enabling faster, more adaptive, and multi-sensory interactions \cite{guo_braincomputer_2008,liu_bi-directional_2019,rutkowski_auditory_2011,an_exploring_2014}. Among all stimulation modalities, visual stimulation typically elicits the strongest and most reliable evoked potentials, and is also the easiest to implement and control in experimental settings.
Nevertheless, the effectiveness of visual stimulation in ERP-BCIs remains highly sensitive to factors such as cue design \cite{guger_comparison_2016} and the spatial density of visual elements \cite{fernandez-rodriguez_influence_2023}, which can significantly affect both performance and perceived fatigue.

Lotte and Jeunet have argued that \ac{bci} training has traditionally emphasized signal processing and algorithmic optimization, while largely overlooking insights from psychology and cognitive science \cite{lotte_flaws_2013,jeunet_why_2016}. This gap has been particularly evident in \ac{mi}-based BCIs, where supervised instructional user training often yields unsatisfactory outcomes. They propose that incorporating principles from \ac{hci} and UX research, along with more diverse and adaptive training strategies, could significantly improve the learning process and overall usability of \ac{bci} systems \cite{lotte_towards_2015}.

Reuderink’s foundational review \cite{reuderink_games_2008} highlights that \ac{bci}-based games do not impair user experience in non-clinical settings and may even enhance it, provided the game design accounts for the inherently lower \ac{itr} of \ac{bci} input. Pioneering studies like Leeb’s wheelchair training simulator \cite{leeb_self-paced_2007} and \textit{Brain Invaders} \cite{congedo_brain_2011} demonstrated how game-based contexts can increase motivation and enable effective online evaluation.

Recently, Glavaš et al. \cite{glavas_evaluation_2022} reported improved BCI control over time with repeated runs of \textit{Zombie Jumper}. In the domain of motor imagery BCIs, gamified training protocols have shown to enhance both user performance and perceived engagement \cite{atilla_improving_2023,atilla_gamification_2024}. 
Another study on attention training for individuals with ADHD \cite{e1966b4104b5466c9b3e481a1837a9c3}, observed comparable performance between gamified and non-gamified conditions, with a marked user preference for the gamified version. 

Despite successful cross-domain adoption, the diversity of use cases means that gold-standard paradigms for game-based research in \ac{bci} remain elusive, and direct comparisons between studies are often limited.

To address this, recent efforts such as the P300 Dynamic Cubes platform \cite{kinney-lang_designing_2020} and its Python-based successor \textit{BCI Essentials} aim to standardize game development pipelines for \ac{bci} research. Nevertheless, building BCI-driven games that are robust, broadly compatible, and efficient for scalable data collection remains a considerable challenge. In particular, ensuring compatibility with consumer-grade EEG hardware and making the systems runnable on standard computing platforms—without requiring high-end processing or lab-grade setups—adds further complexity. These constraints can significantly hinder both reproducibility and accessibility, limiting broader adoption and large-scale data collection efforts in real-world environments.

\section{Methods}\label{methods}
\subsection{Experimental Setup}\label{setup}
\begin{figure*}[ht]
    \centering
    \includegraphics[width=\textwidth]{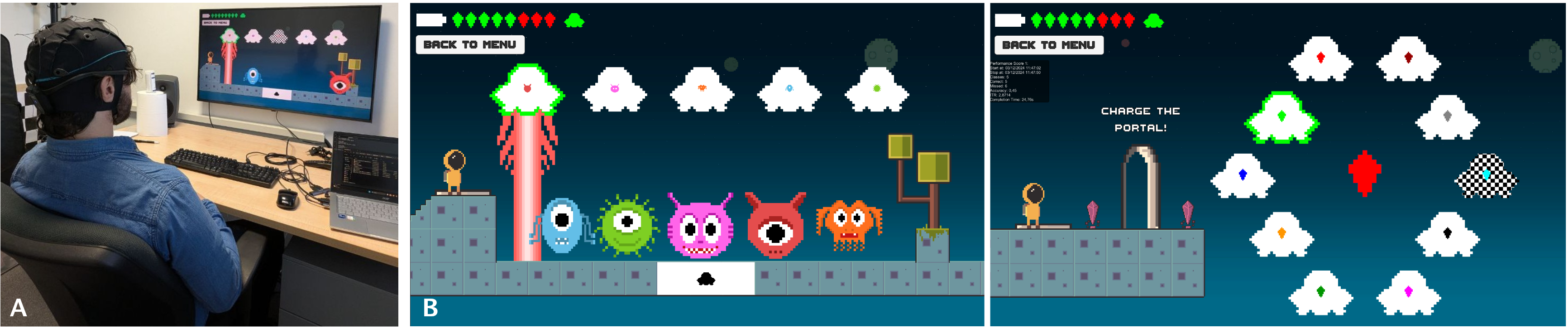}
    \caption{A) A participant taking part into the \textit{BrainForm} experiment. B) Left: an example of the Complex Task; right: an example of the Speller Task. The color of the target is represented by the small symbol in the middle of the alien ships. The green outline provides real-time feedback on the spelled symbol. Targets flash sequentially for 100ms each.}
    \label{fig:tasks}
\end{figure*}

The experiment was conducted at the University of Trento with 22 student participants (10 females; mean age = 21.87, SD = 3.22). All were experienced computer users, though some reported limited familiarity with video games. Only one had prior experience with \acp{bci}, and all were healthy with no self-reported neurological conditions—the only exclusion criterion.
We used the g.tec Unicorn Hybrid Black wearable EEG headset, with conductive gel applied to eight electrodes (Fz, C3, Cz, C4, Pz, PO7, Oz, PO8) following the standardized 10–20 system, and reference/ground electrodes placed on the mastoids (M1, M2).
Participants sat at a desk with a 42" monitor, keyboard, and mouse connected to the researcher’s laptop, which was also mirrored for observation (Fig.\ref{fig:tasks}A). They were instructed to remain seated and minimize movement during trials to ensure signal quality.

\subsection{BrainForm}\label{brainform}
\textit{BrainForm} was developed using Unity 2022.3 and integrates the native SDK for the Unicorn \ac{eeg} device. This package provides essential components for building the \ac{bci} system, including signal quality estimation, a configurable processing pipeline and a proprietary LDA classifier. The game is a 2D platformer that features a tutorial level to introduces players to a simplified version of the experimental tasks, followed by a main level containing the complete task sequence.  The source code and a playable Windows build are freely accessible\footnote{https://github.com/BRomans/BrainForm}.

\subsection{Tasks}
We designed two in-game tasks to evaluate participants’ ability to simultaneously learn game mechanics and develop control over a \ac{bci} system.

The first task, referred to as the \textbf{Complex Task}, is more game-oriented. Players must activate one of five colored laser beams to hit an alien NPC\footnote{Not Playing Character, in gaming jargon.} of the corresponding color. This requires learning how to coordinate timing of \ac{bci} activations with the movements of the NPCs by focusing on the correct flickering target at the right moment. Errors are counted only when the laser fails to hit the matching alien; collisions with incorrectly colored NPCs are not penalized.

The second task, termed the \textbf{Speller Task}, is more BCI-oriented. It consists of a symbol-spelling process embedded in a puzzle narrative, where the player must "unlock" a door by selecting the correct color based on a visual cue. Ten flickering targets represent different colors, and the cue appears at the center of the screen. The sequence of cues is randomized, and an error is recorded when an incorrect target is selected.

Both tasks were refined through a pilot study to ensure balanced difficulty: neither too simplistic nor excessively demanding.
Throughout this paper, repetitions of the Complex Task are labeled as Task A, and repetitions of the Speller Task as Task B, each prefixed by an ordinal number (e.g., \textit{T1A} for the first repetition of Task A, \textit{T3B} for the third repetition of Task B). An in-game example of each is shown in Fig.\ref{fig:tasks}B.

\subsection{Data Collection}\label{data}
We prepared three questionnaires to support the study. An introductory questionnaire was administered at the beginning to collect demographic and info-graphic information. A Session Questionnaire was provided between game-play sessions to assess user comfort, satisfaction, ocular irritation, and to gather qualitative feedback on game aesthetics. Finally, the In-Game module of the \ac{geq} \cite{3d28afa4d1c8412dbe83654f00dcae0f} was used to evaluate participants' user experience in relation to the game-based training tasks. The item \textit{“I was interested in the game's story”} and related aggregate score were omitted, as they were not applicable to \textit{BrainForm} and could have introduced bias into the statistical analysis.
During the experiment, raw \ac{eeg} data, performance metrics and in-game metadata were collected. The \ac{eeg} data collection begins as soon as the device is connected, and includes event triggers for each flickering target. Metrics and metadata are only computed and collected during the tasks onset and include start and end timestamps, the number of classes of the task, the number of correct selections, the number of incorrect selections, the \ac{itr}, task accuracy and total completion time. The full dataset comprises all raw EEG and metadata from the 22 participants who completed the study and is publicly available~\cite{romani_2025_17225966}.
 
\subsection{Experimental Protocol}\label{protocol}
The experiment was structured into two main sessions, separated by a short break and a swap in visual texture.

Prior to the experiment, participants completed a demographic and info-graphic questionnaire, signed an informed consent form, and disclosed any relevant medical conditions. The EEG setup was performed using the Unicorn \ac{eeg} headset with conductive gel. Signal quality and stability were verified using the manufacturer’s diagnostic tools, and the setup was adjusted as needed to ensure high-quality data acquisition. Participants were instructed to minimize movement during recording to reduce motion artifacts.

In the first session, participants were randomly assigned one of two visual textures and initiated the first \ac{bci} calibration procedure, which consisted in focusing on a single flashing target for a total of 60 trials ($\sim$1 minute).

Calibration feedback was color-coded based on cross-validation accuracy: red (<80\%), yellow (80–90\%), and green (>90\%). Calibration was repeated as necessary until at least a yellow-level accuracy was achieved.

Participants then played a tutorial run featuring simplified tasks, followed by the first full practice run. Afterward, they completed a session questionnaire. The second session began with the alternative visual texture, followed by a new calibration, a second practice run, and a final timed challenge. Without removing the equipment, participants completed a second session questionnaire and the \ac{geq}.

At the end of the study, participants were offered the option to play an additional free-play run using their preferred texture. Sixteen participants opted to complete this extra session, which included a new calibration and one additional run without a tutorial. The diagram of the protocol is in Fig.\ref{fig:protocol}.

\subsection{BCI Configuration and Signal Processing}\label{configuration}
The built-in preprocessing pipeline applies a 2nd-order notch filter at both 50Hz and 60Hz to eliminate power-line interference. Additionally, a 6th-order band-pass filter between 0.5Hz and 15Hz is used to suppress movement artifacts and remove irrelevant neural activity outside the frequency range of interest. The data stream is automatically filtered through the pipeline and then segmented between the event onset and 900ms after the event. The parameters of the \ac{bci} were tuned during a prior pilot study.
The calibration length was set to 60 trials, each target flicker was presented for 100ms, and the classifier's regularization coefficient was set to 1. To reduce the likelihood of false activations, the confidence threshold for classifier output was increased to 0.95, and a minimum uninterrupted focus duration of 500ms was enforced to activate a target.

\section{Results}\label{results}

\subsection{Calibration Attempts}\label{calibration}

On average, participants required $2.64 \pm 1.33$ calibration attempts during the first session and $2.68 \pm 1.73$ during the second. Among those who chose to complete the optional free-play session, the number of calibration attempts averaged $2.47 \pm 1.23$ in the first session, decreased to $2.29 \pm 1.65$ in the second, and then to $1.94 \pm 1.60$ in the final session, but the trend is not significant.

\begin{table}[t!]
\centering
\resizebox{1.0\textwidth}{!}{
\begin{tabular}{l|rrrr|rrrr|rrrr|rrr}
\toprule
\textbf{A Run} 
& \multicolumn{4}{c|}{\textbf{Accuracy (Median) I}} 
& \multicolumn{4}{c|}{\textbf{Time(s) (Median) II}} 
& \multicolumn{4}{c|}{\textbf{ITR (Mean) III}} 
& \multicolumn{3}{c|}{\textbf{Decision}} \\
 & MED & MAD & $p_{\text{equal}}$ & $p_{\text{sm}}$ 
 & MED & MAD & $p_{\text{equal}}$ & $p_{\text{sm}}$ 
 & M & SD & $p_{\text{equal}}$ & $p_{\text{sm}}$ 
 & A & T & I \\
\midrule
t1a 
& \textbf{0.714} & 0.138 & 0.000 & 0.924 
& 46.15 & 6.900 & 0.000 & 0.339 
& \textbf{7.99} & 5.315 & 0.000 & 0.952 
& ↓ & i & ↓ \\
t2a $^I$ $^{II}$ $^{III}$ 
& 0.833 & 0.167 & --- & ---
& 46.80 & 14.100 & --- & ---
& 10.76 & 7.242 & ---  & ---
& – & – & – \\
t3a 
& \textbf{0.714} & 0.119 & 0.000 & 0.985 
& 46.45 & 8.350 & 0.000 & 0.334 
& \textbf{6.94} & 5.181 & 0.000 & 0.988 
& ↓ & i & ↓ \\
\toprule
\textbf{B Run} 
& \multicolumn{4}{c|}{\textbf{Accuracy (Median) I}} 
& \multicolumn{4}{c|}{\textbf{Time(s) (Median) II}} 
& \multicolumn{4}{c|}{\textbf{ITR (Mean) III}} 
& \multicolumn{3}{c|}{\textbf{Decision}} \\
 & MED & MAD & $p_{\text{equal}}$ & $p_{\text{sm}}$ 
 & MED & MAD & $p_{\text{equal}}$ & $p_{\text{sm}}$ 
 & M & SD & $p_{\text{equal}}$ & $p_{\text{sm}}$ 
 & A & T & I \\
\midrule
t1b $^{II}$
& \textbf{0.773} & 0.106 & 0.001 & 0.938
& 72.80 & 12.250 & --- & ---
& \textbf{15.77} & 8.006 & 0.000 & 0.983 
& ↓ & – & ↓ \\
t2b 
& \textbf{0.769} & 0.121 & 0.000 & 0.939 
& 65.75 & 16.600 & 0.000 & 0.884 
& 18.89 & 9.024 & 0.000 & 0.918 
& ↓ & i & i \\
t3b $^{I}$ $^{III}$
& 0.833 & 0.076 & --- & ---
& \textbf{63.00} & 14.150 & 0.000 & 0.973
& 21.95 & 9.669 & --- & ---
& – & ↓ & – \\
\bottomrule
\end{tabular}
}
\caption{Compact summary of task metrics and Bayesian decision outcomes for Accuracy, Task Time, and ITR across three game-play runs.
\\[0.5ex]
\textbf{M}: Mean; \textbf{MED}: Median; 
\textbf{SD}: Standard Deviation; \textbf{MAD}: Median Absolute Deviation; 
$p_{\text{equal}}$: probability of no difference; 
$p_{\text{sm}}$: probability the current value is smaller than the reference; 
\textbf{Decision (A | T | I)}: ↓ = significantly lower, i = inconclusive, -- = reference row.
Values in bold reflect the significant change reported in the decision column.
Rows marked with roman numerals (I) indicate for which column they were automatically selected as reference by \texttt{autorank}.}
\label{tbl:stat_results_3runs}
\end{table}
\subsection{User Questionnaires}\label{eyes}
\begin{figure}[h]
    \centering
    \begin{subfigure}[b]{0.48\textwidth}
        \centering
        \includegraphics[width=\textwidth]{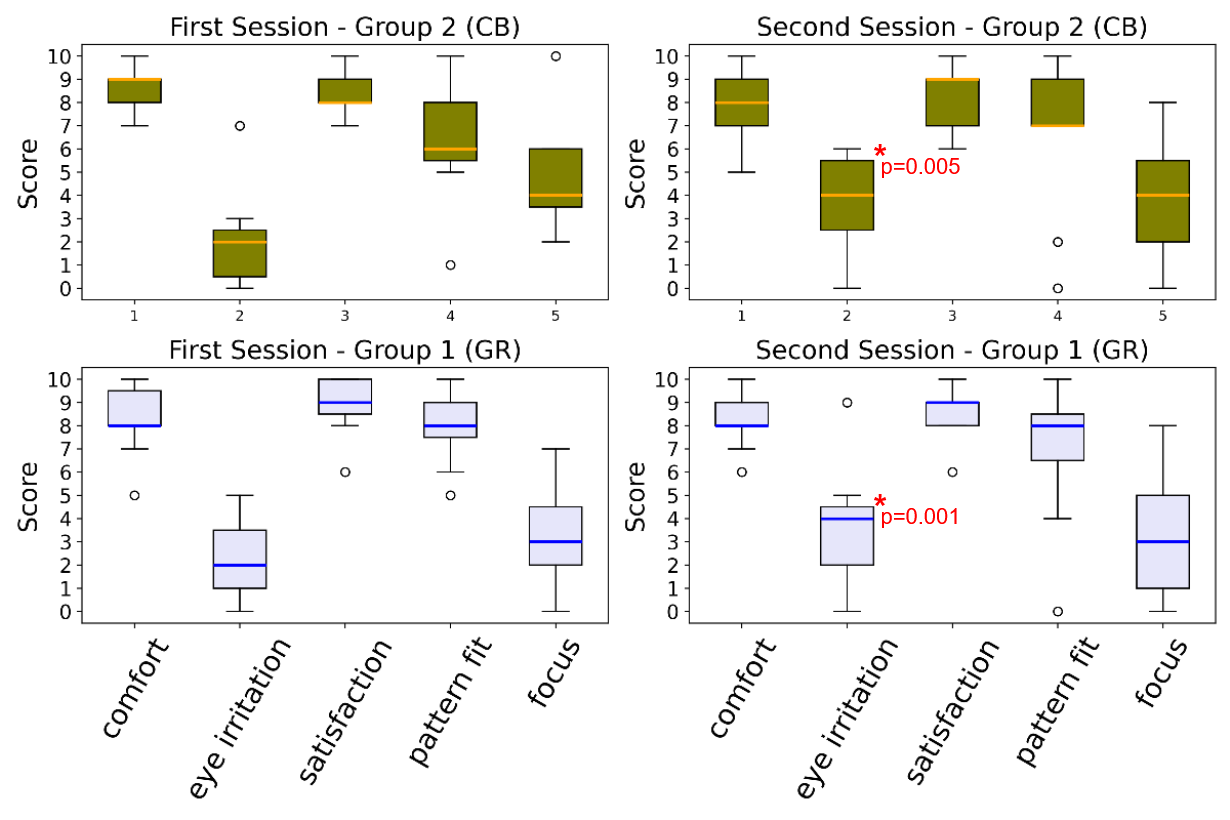}
        \caption{Session questionnaire results for both groups. Ocular irritation significantly increases after the second run; $*$ marks $p_{equal}$.}
        \label{fig:session_questionnaire}
    \end{subfigure}
        \hfill
    \begin{subfigure}[b]{0.48\textwidth}
        \centering
        \includegraphics[width=\textwidth]{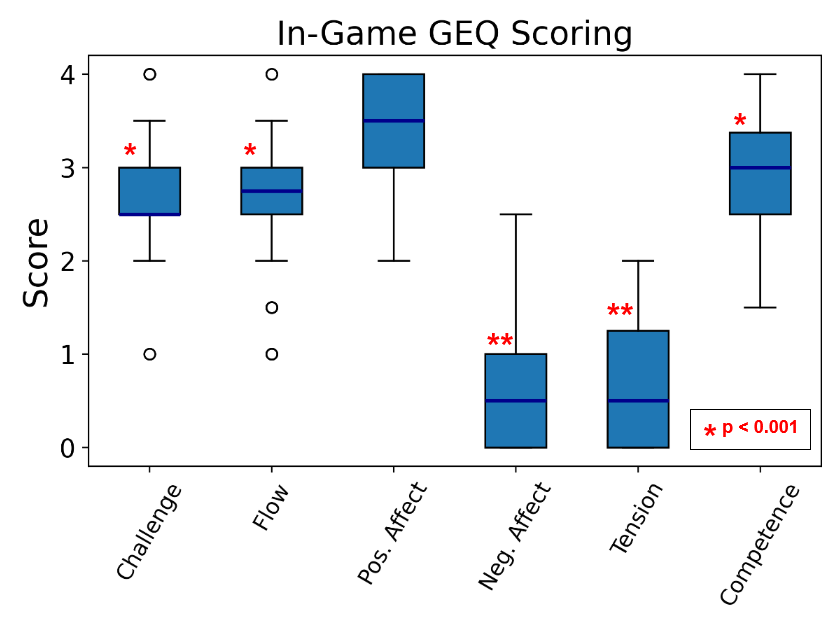}
        \caption{Aggregate scores of the In-Game GEQ module. A single $*$ marks $p_{equal}$ to \textit{Pos. Affect}. A double $**$ marks $p_{equal}$ to all the other items and \textit{Pos. Affect}.}
        \label{fig:GEQ}
    \end{subfigure}

    \caption{Results from post-session questionnaires: (a) Session questionnaires outcome  and (b) In-Game GEQ scores.}
    \label{fig:GEQ_and_session}
\end{figure}

Subjects expressed their preference in terms of visual texture as follow: 8 grain-like, 10 checkerboard, 4 no preference.
Bayesian signed-rank tests did not reveal any significant differences between the two visual texture groups in each individual game-play session. However, within-group comparisons showed notable changes between the first and second sessions. Specifically, \textit{Ocular Irritation} significantly increased in Group 1 (GR) from the first session ($M = 2.091 \pm 1.558$, $SD = 1.700$) to the second session ($M = 3.545 \pm 2.220$, $SD = 2.423$). A similar significanttrend was observed in Group 2 (CB). When aggregating data across both groups, \textit{Ocular Irritation} rose from $M = 2.045 \pm 1.055$, $SD = 1.812$ in the first session to $M = 3.636 \pm 1.252$, $SD = 2.150$ in the second session as shown in Fig.\ref{fig:session_questionnaire}.

The \ac{geq} results (Fig. \ref{fig:GEQ}) indicate a generally positive reception of the gamified experience. Aggregate scores for \textit{Competence}, \textit{Sensory and Imaginative Immersion}, \textit{Flow}, and \textit{Positive Affect} were all significantly higher than\textit{Tension} and \textit{Negative Affect}.

\begin{table}[t!]
\centering
\resizebox{1.0\textwidth}{!}{
\begin{tabular}{l|rrrr|rrrr|rrrr|rrr}
\toprule
\textbf{A Run} 
& \multicolumn{4}{c|}{\textbf{Accuracy (Mean) I}} 
& \multicolumn{4}{c|}{\textbf{Time(s) (Median) II}} 
& \multicolumn{4}{c|}{\textbf{ITR (Mean) III}} 
& \multicolumn{3}{c|}{\textbf{Decision}} \\
 & M & SD & $p_{\text{equal}}$ & $p_{\text{sm}}$ 
 & MED & MAD & $p_{\text{equal}}$ & $p_{\text{sm}}$ 
 & M & SD & $p_{\text{equal}}$ & $p_{\text{sm}}$ 
 & A & T & I \\
\midrule
t1a     
& 0.750 & 0.180 & 0.000 & 0.790 
& 42.25 & 3.300 & 0.000 & 0.935 
& 8.20 & 5.73 & 0.000 & 0.725 
& i & i & i \\
t2a$^I$ $^{III}$ 
& 0.789 & 0.217 & -- & -- 
& \textbf{44.50} & 12.150 & 0.000 & 0.982 
& 10.57 & 7.888 & -- & -- 
& -- & ↓ & -- \\
t3a     
& \textbf{0.654} & 0.193 & 0.001 & 0.982 
& 46.00 & 5.800 & 0.005 & 0.556 
& \textbf{5.65} & 4.658 & 0.000 & 0.987 
& ↓ & i & ↓ \\
t4a$^{II}$      
& 0.696 & 0.183 & 0.006 & 0.947 
& 47.10 & 8.650 & -- & -- 
& \textbf{6.31} & 4.504 & 0.000 & 0.987 
& i & -- & ↓ \\
\toprule
\textbf{B Run} 
& \multicolumn{4}{c|}{\textbf{Accuracy (Median) I}} 
& \multicolumn{4}{c|}{\textbf{Time(s) (Median) II}} 
& \multicolumn{4}{c|}{\textbf{ITR (Mean) III}} 
& \multicolumn{3}{c|}{\textbf{Decision}} \\
 & MED & MAD & $p_{\text{equal}}$ & $p_{\text{sm}}$ 
 & MED & MAD & $p_{\text{equal}}$ & $p_{\text{sm}}$ 
 & M & SD & $p_{\text{equal}}$ & $p_{\text{sm}}$ 
 & A & T & I  \\
\midrule
t1b$^{II}$    
& \textbf{0.714} & 0.119 & 0.002 & 0.962 
& 70.90 & 11.850 & -- & -- 
& \textbf{14.67} & 8.050 & 0.000 & 1.000 
& ↓ & -- & ↓ \\
t2b     
& 0.801 & 0.108 & 0.001 & 0.793 
& \textbf{57.25} & 13.000 & 0.000 & 0.992 
& 21.28 & 9.263 & 0.000 & 0.692 
& i & ↓ & i \\
t3b$^I$ 
& 0.833 & 0.076 & -- & -- 
& \textbf{57.05} & 11.300 & 0.000 & 0.989 
& 23.70 & 10.662 & 0.000 & 0.479 
& -- & ↓ & i \\
t4b$^{III}$       
& 0.833 & 0.070 & 0.001 & 0.666 
& \textbf{55.70} & 10.500 & 0.000 & 1.000 
& 23.70 & 9.617 & -- & -- 
& i & ↓ & -- \\
\bottomrule
\end{tabular}
}
\caption{
Compact summary of task metrics and Bayesian decision outcomes for Accuracy, Task Time, and ITR across four game-play runs. Same legend as Table 1 applies.
}
\label{tbl:stat_results_4runs}
\end{table}

\subsection{Online Performance}\label{online-metrics}
Since no statistically significant differences were found between groups in the preliminary analysis, all participants were pooled into a single population for the within-trial analysis using Bayesian signed-rank tests. To simplify the complexity of statistical comparisons, the analysis was partially automated using \texttt{autorank} \cite{Herbold2020}.
Tutorial tasks were excluded, resulting in three runs per task, with an optional fourth run completed by a subset of volunteers.

The results of the online performance metrics are presented in two stages:
(1) including all participants, and
(2) including only those who volunteered for the optional fourth run.

For each task, results are reported in the following order: \textit{Task Accuracy}, \textit{Task Time}, and \textit{Information Transfer Rate (ITR)}.
Note that all descriptive statistics (mean or median) are reported as selected automatically by the \texttt{autorank} procedure, based on the data distribution and statistical assumptions.

\subsubsection{Analysis Of The Entire Population}\hfill\\
\textit{Complex Task (A)} —  
\textbf{Task Accuracy:} T3A is significantly lower than T2A following the time challenge;  
\textbf{Task Time:} no significant difference observed;  
\textbf{ITR:} T3A is significantly lower than T2A, consistent with the accuracy trend.

\textit{Speller Task (B)} —  
\textbf{Task Accuracy:} no significant difference observed;  
\textbf{Task Time:} T1B is significantly higher than T3B;  
\textbf{ITR:} both T2B and T3B are significantly higher than T1B. Notably, despite the time challenge, T3B showed the highest values for both \textit{Task Accuracy} and \textit{ITR}. All values and statistical comparisons are summarized in Table~\ref{tbl:stat_results_3runs}.

\subsubsection{Analysis of Optional Run}\hfill\\

\textit{Complex Task (A)} —  
\textbf{Task Accuracy:} T3A is significantly lower than both T2A and T1A, consistent with the previous analysis;  
\textbf{Task Time:} T4A (free-play) is significantly longer than T2A;  
\textbf{ITR:} T3A is significantly lower than T2A and T1A. T4A is also significantly lower than T2A, following the same trend as task duration.

\textit{Speller Task (B)} —  
\textbf{Task Accuracy:} T2B, T3B, and T4B are all significantly higher than T1B;  
\textbf{Task Time:} T1B is significantly higher than T2B, T3B, and T4B;  
\textbf{ITR:} T2B, T3B, and T4B all show significantly higher central tendency than T1B, in line with the improvements in both accuracy and speed. 

\begin{figure}[h]
    \centering
    \includegraphics[width=\columnwidth]{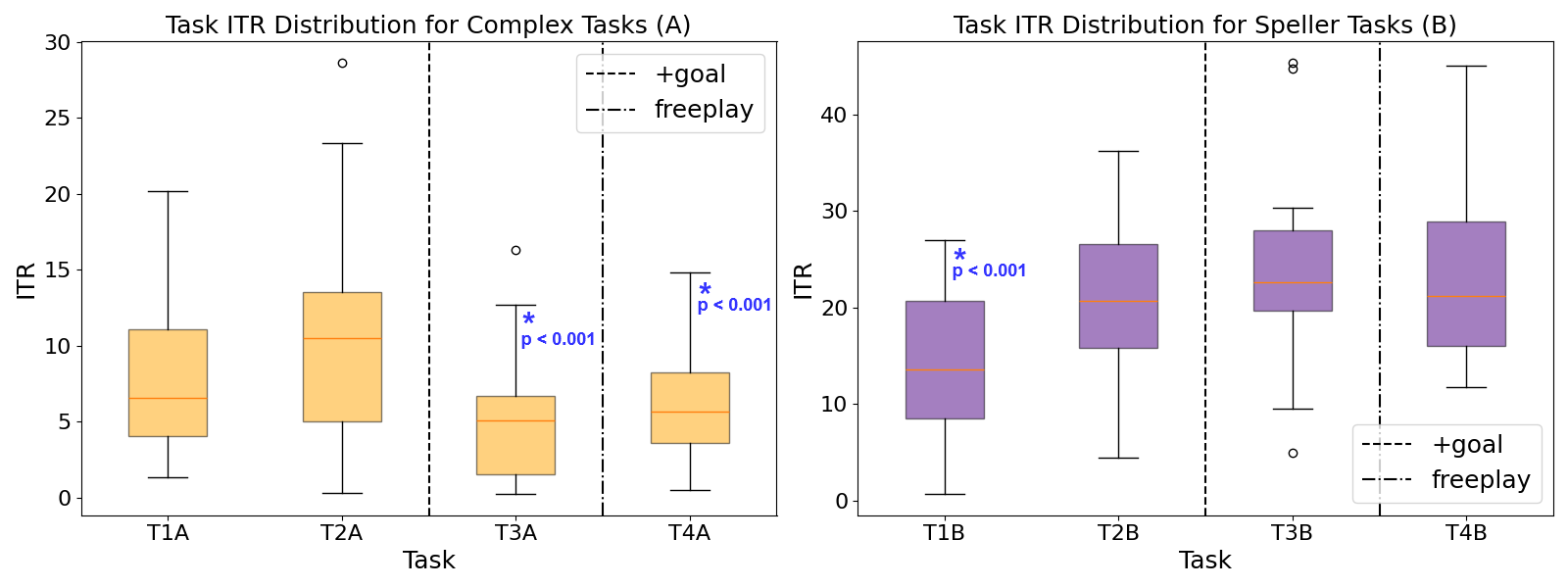}
    \caption{Trends of the ITR for repetition of Complex Task and repetitions of the Speller Task; $*$ marks $p_{smaller}$ than the reference run, as per Table \ref{tbl:stat_results_4runs}.} 
    \label{fig:box_itr}
\end{figure}

All values and statistical comparisons are summarized in Table~\ref{tbl:stat_results_4runs} and \ac{itr} trends are visualized in Fig. \ref{fig:box_itr}.

\section{Discussion}\label{discussion}

\textbf{Questionnaires}. Concerning \textbf{RQ1}, the session questionnaires did not indicate any perceived difference nor strong preference between the two visual textures of the stimuli. This can be interpreted as a positive outcome, suggesting that deviations from the standard checkerboard texture are acceptable from a user perspective. However, both groups reported a noticeable increase in ocular irritation following the second phase of the experiment, in line with the aforementioned limitations.
Further investigation should delve into alternative pattern designs, including variations in color combinations and spatial densities, as suggested by multiple participants.

The GEQ, by contrast, proved effective in capturing participants' appreciation for both the game design and the overall experimental session. Future user studies using \textit{BrainForm} may adopt alternative assumptions to investigate gender and age-related differences, both often overlooked in \ac{bci} and game-based research due to the persistent challenge of recruiting balanced participant populations. This problem is further compounded by individual differences in gaming experience, which can significantly influence engagement, motivation, competitiveness and performance.

In line with observations by Reuderink, Lotte and Jeunet \cite{reuderink_games_2008,lotte_flaws_2013,jeunet_why_2016}, our results confirm that user-centered, game-based training can improve experience without compromising performance. 
Compared to systems like \textit{Brain Invaders} \cite{congedo_brain_2011}, \textit{BrainForm} emphasizes modularity, consumer hardware, and in-game metadata collection, enabling more scalable and repeatable data collection.

\textbf{Online Metrics}. Regarding the online metrics and in relation to \textbf{RQ2}, we did not find evidence that one visual texture improved or impaired user performance over the other. This may be attributed to the limited scope of this exploratory part, as only two black-and-white patterns were evaluated. 
Addressing \textbf{RQ3} and \textbf{RQ4}, our findings clearly suggest the following: (1) participants demonstrated learning effects, improving both their game-play and \ac{bci} control skills after the initial run; (2) introducing a time challenge had a measurable impact on the performance of the A tasks, which requires precise timing for successful completion; (3) the time pressure did not impair users' ability to control the \ac{bci}, even though the number of flickering targets was doubled in B tasks compared to A tasks; (4) in run T3B, participants achieved the highest average \textit{Task Accuracy} among the three core runs, along with a significantly lower \textit{Task Time} compared to the first run and a higher, but not statistically significant, \textit{\ac{itr}}.
In addition, in run T4A, the increased \textit{Task Time} accompanied by improved \textit{Task Accuracy} suggests a user-driven adaptation, potentially as a recovery strategy following the previous suboptimal performance. 

Nevertheless, we believe we were able, to some extent, to disentangle the processes of learning game mechanics and acquiring \ac{bci} control skills. This was demonstrated by introducing a time constraint, that revealed that errors increased when precise timing was required, whereas \ac{bci} control performance, measured via symbol-spelling, consistently improved across runs. Furthermore, the observed learning effects and subjective engagement scores support the findings of Glavaš et al. \cite{glavas_evaluation_2022} on repeated-session adaptation, while offering a lightweight and extensible alternative for varied experimental designs.
       
\textbf{Ergonomics}. Multiple failed calibration attempts remained a source of frustration for participants and a drawback for the user experience. This is not solely imputable to the user’s ability to focus or perform the task, but also on the responsiveness and adaptability of the classifier. In some cases, sessions lasting up to 10 minutes not only delayed the start of the experiment, but also increased visual fatigue before the tasks began. Although essential for achieving sufficient \ac{bci} accuracy, the procedure often required reapplying conductive gel and repositioning electrodes. In only two cases, participants were unable to continue and had to be excluded. While such a burden may be tolerated under experimental conditions, it would pose a major obstacle to the adoption of consumer-grade \ac{bci} systems. More robust calibration strategies should account for both user variability and system limitations.

Visual fatigue is a known factor elicited by and detrimental for ERP-based \acp{bci}, yet remains largely unexplored in the literature. In addition to variations in spatial pattern density, previous studies have investigated alternative strategies to mitigate visual strain, such as (1) restricting participants’ visual field using a binocular aperture~\cite{Kirasirova2020-fy}, and (2) reducing the color contrast of the visual stimuli~\cite{8008365}. While the first approach seems impractical, the second appears more feasible compared to our method, which employs stimuli at the extreme ends of the contrast scale (black \& white).

\textbf{Limitations}. The study setup prioritized simplicity to allow multiple calibration sessions and \ac{bci} tasks within a one-hour timeframe. As a result, the findings do not permit conclusions about long-term acquisition or retention of \ac{bci} control skills. Another limitation arises from the fixed task order imposed by the game design, which may have introduced fatigue effects in later runs. Participant selection involved minimal prerequisites, resulting in variability in prior gaming experience that could have affected learning curves or in-game strategies. The young age and relatively small size of the sample may also have biased participants’ reception of the video game as a training medium. Finally, the study focused solely on spatial patterns and a single contrast level for textures, limiting the scope of visual factor assessment.

\section{Conclusions}\label{conclusions}
We introduced BrainForm, a serious neurogame designed for scalable \ac{bci} user studies and data collection. Our study investigated the effects of two visual stimulation textures and examined the progression of \ac{bci} control skills across repeated sessions. We invite the community to build upon our open-source code and dataset. From an interaction and design standpoint, it is essential to deepen our understanding of the learning dynamics and affordances of \ac{bci} technology to create smoother and more engaging user experiences. 
Our future work will focus on enhancing generalization using lightweight deep neural networks, exploring subject-specific conditioning, and deploying the system on resource-constrained hardware.

\bibliographystyle{splncs04}
\bibliography{Brainform}

\end{document}